\documentstyle[eqsecnum,aps,prl,epsf]{revtex}
\title{
Extended bound states and resonances of two fermions on a periodic lattice
}
\author{A. S. Blaer$^{a}$, H. C. Ren$^{b}$, and O. Tchernyshyov$^{a}$}
\address{$^{a}$Department of Physics, Columbia University, 
New York, NY 10027}
\address{$^{b}$Department of Physics, Rockefeller University, New York, 
NY 10021}
\date{August 7, 1996} 
\draft
\preprint{CU-TP-771, RU96-6-B, supr-con/9608003}
\begin{document}
\bibliographystyle{unsrt}

\maketitle
\begin{abstract}
The high-$T_c$ cuprates are possible candidates for 
$d$-wave superconductivity, with the Cooper
pair wave function belonging to a non-trivial irreducible
representation of the lattice point group.  We argue that 
this $d$-wave symmetry is related to a special form of the
fermionic kinetic energy and does not require any
novel pairing mechanism.  In this context, we present 
a detailed study of the bound states and resonances formed by two 
lattice fermions interacting via a non-retarded potential that is 
attractive for nearest neighbors but repulsive for other 
relative positions.  In the case of strong binding, a pair formed by 
fermions on adjacent lattice sites can have a small effective mass, 
thereby implying a high condensation temperature.  For a weakly bound 
state, a pair with non-trivial symmetry tends to be smaller in size 
than an $s$-wave pair.  These and other findings are 
discussed in connection with the properties of high-$T_c$ cuprate
superconductors. 
\end{abstract}

\section{Introduction}\label{intro}

Recently, through angle-resolved photoemission spectroscopy \cite{Shen} 
and the tri-crystal Josephson effect \cite{Tsuei}, a substantial 
$d$-wave component has been 
detected in the order parameter of the cuprate superconductors. 
For tetragonal crystals, such as Tl-based cuprates in which the 
square symmetry of the $\rm CuO_2$ plane is perfect, the order 
parameter appears to be entirely $d$-wave.  In
other superconducting cuprates, with orthorhombic distortions,
the order parameter contains a large
fraction of $d$-wave. These observations have prompted theorists
to look for exotic mechanisms that could lead to a non-trivial 
symmetry of the order parameter.  It may, however, be true  
that it is simply the fermionic kinetic energy
that leads to $d$-wave superconductivity.  
An exactly solvable problem with two fermions would provide a strong
argument for this possibility.  

Pines {\em et al.} \cite{Pines} have proposed a one-band model for cuprate 
superconductors in 
which quasiparticles hop between copper sites with amplitude $t_1<0$ along 
the sides of square cells and with amplitude $t_2>0$ along the diagonals.  
The interaction 
between quasiparticles is mediated by magnon exchange and, in the static limit,
can be thought of as a repulsive potential for two quasiparticles on the 
same antiferromagnetic sublattice and as an attractive 
potential otherwise. In the present paper,
we demonstrate that a positive $t_2$ is actually a necessary condition 
for a two-fermion system to have a $d$-wave ground state. In cuprate
compounds, a reversed (positive) sign of $t_2$ is indeed expected if 
doped 
holes occupy $d_{x^2-y^2}$ orbitals of the copper ions. Such orbitals have 
four lobes along the principal axes of the $\rm CuO$ planes, with a positive 
amplitude along one axis and a negative amplitude along the other. For 
adjacent copper 
ions the overlapping lobes have the same sign, while for next-nearest 
neighbors positive lobes overlap with negative ones. Therefore, a reversed 
sign of $t_2$ occurs.

Measurements of the upper critical magnetic field have revealed a very
short coherence length in high-$T_c$ cuprates. As we shall see, 
for a $d$-wave pair, the average distance between the
two fermions is on the order of the lattice spacing, 
even for weak binding. Therefore, these 
pairs are very localized and may well undergo a Bose-Einstein 
condensation. 
The resulting transition temperature $T_{\rm BE}$ decreases with a 
growing effective boson mass. For a point-like boson on a 
lattice  (strong on-site binding), the mass is large because the particle 
can only move via virtual fermionic states in which the pairing 
is broken \cite{Nozieres}. As 
we shall show, this is not the case for an extended boson such as the 
$d$-wave pair discussed in this article, and the suppression 
of $T_{\rm BE}$ disappears. 

Phenomenological models with resonance bosons have been 
proposed by Friedberg and Lee \cite{FL89} and by 
Micnas, Ranninger, and Robaszkiewicz \cite{Micnas90}. 
The $s$-wave description given in \cite{FL89} can be extended to include 
the possibility of resonances having non-trivial symmetry. In the 
present paper, we discuss the formation of a two-fermion resonance 
($s$-wave or $d$-wave) and show that such a resonance boson can propagate 
long distances before decaying.

The present article is organized as follows: In Section \ref{formalism}, 
we describe
the fermionic Hamiltonian and present a theorem concerning 
the symmetry of the two-fermion ground state. Deep bound states, 
shallow bound states, and resonances are analyzed in 
Sections \ref{deep}, \ref{shallow}, and \ref{resonance}. The discussion in 
these sections is mostly restricted to the case of a square lattice, 
with the exception 
of Section \ref{fcc}, in which we deal with a face-centered cubic lattice 
as in fullerene superconductors. Concluding remarks are made in 
Section \ref{conclusion}.

\section{Two Fermions on a Periodic Lattice}\label{formalism}
\subsection{The Hamiltonian and its properties}
\label{basic notations}

We consider a periodic lattice of ${\cal N}$ sites with symmetry under 
inversion at each site. The fermionic 
kinetic energy is represented by a sum of "hopping" terms, 
\begin{equation}
\label{kinetic energy}
H_0 =\sum_{s,{\bf r},{\bf r}'} 
t({\bf r}-{\bf r}') a_s^\dagger({\bf r}') a_s({\bf r}),
\end{equation}
with $s=\ \uparrow$ (spin up) and $\downarrow$ (spin down). The interaction 
between two particles is described in terms 
of a potential energy that is a function of the relative separation and the 
total spin of a pair:
\begin{eqnarray}
H_1= \sum_{{\bf r},{\bf r}'}[ 
&& g_S({\bf r}-{\bf r}')\ {\cal S}({\bf r},{\bf r}')^\dagger{\cal S}({\bf r},
{\bf r}')
\nonumber\\
\label{interaction}
{} +&& g_T({\bf r}-{\bf r}')\ \sum_\mu {\cal T}_\mu^\dagger({\bf r},{\bf r}')
{\cal T}_\mu({\bf r},{\bf r}')],
\end{eqnarray}
where $g_S({\bf r})$ and $g_T({\bf r})$ represent interaction energies 
in singlet and 
triplet states with a relative separation ${\bf r}$ and where
${\cal S}({\bf r},{\bf r}')$ and ${\cal T}_\mu({\bf r},{\bf r}')$ are the 
destruction operators 
for spin-singlet and spin-triplet states on sites ${\bf r}$ and ${\bf r}'$, 
that is,
\begin{mathletters}
\begin{eqnarray}
{\cal S}({\bf r},{\bf r}') &=& 
\frac{a_\uparrow({\bf r}) a_\downarrow({\bf r}') 
+ a_\uparrow({\bf r}') a_\downarrow({\bf r})}
{2},\\ 
{\cal T}_0({\bf r},{\bf r}') &=&
\frac{a_\uparrow({\bf r}) a_\downarrow({\bf r}') 
- a_\uparrow({\bf r}') a_\downarrow({\bf r})}
{2}, \\ 
{\cal T}_+({\bf r},{\bf r}') &=& 
\frac{a_\uparrow({\bf r}) a_\uparrow({\bf r}')}
{\sqrt{2}}, \\
{\cal T}_-({\bf r},{\bf r}') &=& 
\frac{a_\downarrow({\bf r}) a_\downarrow({\bf r}')}
{\sqrt{2}}.
\end{eqnarray}
\end{mathletters}
The particular normalization of these operators has been chosen for later 
convenience. In this paper we are mainly concerned with 
singlet states.  Therefore, the triplet part of the interaction will  
not contribute, and the subscript in the interaction energy $g_S$ will 
usually be suppressed.

It follows from (\ref{kinetic energy}) that the one-fermion energy 
spectrum is 
\begin{equation}
\epsilon_{\bf p} = \sum_{\bf r} t({\bf r}) \cos{({\bf p}\!\cdot\!{\bf r})}
\end{equation}
with ${\bf p}$ the Bloch momentum. The corresponding effective mass in $D$ 
dimensions is given by
\begin{equation}
\label{fermion mass}
\frac{1}{m} = -\frac{1}{D}\sum_{\bf r} t({\bf r})r^2 
\end{equation}
for the case in which the inverse mass-matrix $(m^{-1})_{ij}=\delta_{ij}/m$.
In this paper, we only consider the case for which $m>0$.
 
A convenient basis set in the Hilbert space of two fermions 
consists of states with total Bloch momentum ${\bf P}$, relative separation 
${\bf r}$, and spin projections $s_1$ and $s_2$:  
\begin{equation}
\label{Prss}
|{\bf P},{\bf r},s_1,s_2\rangle = \frac{1}{\sqrt{\cal N}} 
\sum_{{\bf r}'}e^{i{\bf P}\cdot({\bf r}'+{\bf r}/2)}\ 
a_{s_1}^\dagger({\bf r}') a_{s_2}^\dagger({\bf r}'+{\bf r}) |0\rangle.
\end{equation}
Applying the kinetic energy operator (\ref{kinetic energy}) to state 
(\ref{Prss}), we obtain 
\begin{equation}
\label{KE eigen}
H_0|{\bf P},{\bf r},s_1,s_2\rangle = 
\sum_{{\bf r}'}2t({\bf r}')\cos{({\bf P}\!\!\cdot\!{\bf r}'/2)}
|{\bf P},{\bf r}+{\bf r}',s_1,s_2\rangle. 
\end{equation}
The problem of two fermions is therefore reduced to that of a single 
particle hopping 
over ${\bf r}$ sites (in the relative position space) with an amplitude 
$2t({\bf r})\cos{({\bf P}\!\!\cdot\!{\bf r}/2)}$. Eigenvectors of $H_0$ are 
states with total momentum ${\bf P}$ and definite relative momentum ${\bf p}$,
\begin{equation}
\label{Ppss}
|{\bf P},{\bf p},s_1,s_2\rangle =
a^\dagger_{({\bf P}/2)-{\bf p},s_1}\ a^\dagger_{({\bf P}/2)+{\bf p},s_2}\ 
|0\rangle,
\end{equation}
with eigenvalues
\begin{eqnarray}
\label{E(p)}
E_{{\bf P},{\bf p}} 
&=& \epsilon_{({\bf P}/2)+{\bf p}} + \epsilon_{({\bf P}/2)-{\bf p}} 
\nonumber \\
&=& 2\sum_{\bf r} t({\bf r})\cos{({\bf P}\!\!\cdot\!{\bf r}/2)}
\cos{({\bf p}\!\cdot\!{\bf r})}.
\end{eqnarray}
The operators $a_{{\bf p},s}$ and $a_s({\bf r})$ are related by a Fourier 
transformation, \begin{equation}
\label{Fourier}
a_{{\bf p},s} = \frac{1}{\sqrt{\cal N}} 
\sum_{\bf r} e^{-i{\bf p}\cdot{\bf r}} a_s({\bf r}).
\end{equation}
 
The projection of any two-fermion state on the state (\ref{Prss}) is the 
wave function in relative position space:
\begin{equation}
\label{psi defined}
\psi({\bf r}) = \langle{\bf P},{\bf r},s_1,s_2|{\bf P},\psi,s_1,s_2\rangle. 
\end{equation}
Plane waves, 
\begin{equation}
\label{plane waves}
\frac{1}{\sqrt{\cal N}}\psi({\bf r}|{\bf p}) \equiv 
\langle{\bf P},{\bf r},s_1,s_2|{\bf P},{\bf p},s_1,s_2\rangle 
=\frac{1}{\sqrt{\cal N}} e^{i{\bf p}\cdot{\bf r}}, 
\end{equation}
form a complete orthonormal set of functions:
\begin{eqnarray}
\label{complete}
\frac{1}{{\cal N}}\sum_{\bf p} \psi^*({\bf r}'|{\bf p})\psi({\bf r}|{\bf p}) 
= \delta_{{\bf r}'{\bf r}},\\
\label{orthonormal}
\frac{1}{{\cal N}}\sum_{\bf r} \psi({\bf r}|{\bf p}')\psi^*({\bf r}|{\bf p}) 
= \delta_{{\bf p}'{\bf p}}.
\end{eqnarray}
For singlet states, the two-fermion wave function must be symmetric 
under the transformation ${\bf r}\rightarrow -{\bf r}$:
\begin{equation}
\label{psi symmetrized}
\frac{1}{\sqrt{\cal N}}\psi({\bf r}|{\bf p}) = 
\langle{\bf P},{\bf r}|{\bf P},{\bf p}\rangle 
 = \sqrt{\frac{2}{{\cal N}}}\cos{({\bf p}\!\cdot\!{\bf r})}
\end{equation}
for ${\bf r}\neq 0$.
For two fermions on the same site (${\bf r}=0$), $\psi(0|{\bf p}) = 1$ 
is already 
symmetric.
Using the symmetrized wave functions (\ref{psi symmetrized}), we can write 
the interaction matrix elements of $H_1$ as 
\begin{equation}
\label{matrix elements}
\langle{\bf P},{\bf q}|H_1|{\bf P},{\bf p}\rangle = 
\frac{1}{{\cal N}}
\sum_{{\bf r}}\psi^*({\bf r}|{\bf q})\,g({\bf r})\,\psi({\bf r}|{\bf p}). 
\end{equation}
Sites ${\bf r}$ and $-{\bf r}$ are now considered equivalent and only one of 
them enters the   sum.

\subsection{Schr\"odinger equation}
\label{Schrodinger equation}

To solve the Schr\"odinger equation,
\begin{equation}
\label{Schrodinger}
(H_0+H_1)|{\bf P},\psi\rangle =E|{\bf P},\psi\rangle,
\end{equation}
for a two-particle bound state, we write 
$|{\bf P},\psi\rangle$ as a sum over states having definite relative momenta:
\begin{equation}
\label{sum of plane waves}
|{\bf P},\psi\rangle = \sum_{\bf p} F({\bf p})|{\bf P},{\bf p}\rangle. 
\end{equation}
The Schr\"odinger equation is then transformed into an integral equation 
for $F({\bf p})$:
\begin{eqnarray}
\label{integral equation}
(E-E_{{\bf P},{\bf p}})F({\bf p}) 
&=& \sum_{\bf q} F({\bf q})
\langle {\bf P},{\bf p}|H_1|{\bf P},{\bf q}\rangle \\
&=& \frac{1}{{\cal N}}\sum_{\bf r} \psi^*({\bf r}|{\bf p}) g({\bf r}) 
\sum_{\bf q} F({\bf q})\psi({\bf r}|{\bf q})
\nonumber.
\end{eqnarray}
Recalling that $\psi({\bf r}) = 
\sum_{\bf q} F({\bf q}) \psi({\bf r}|{\bf q})$, we obtain the following 
equation for the wave function:
\begin{equation}
\label{psi = G g psi}
\psi({\bf r}) = 
\sum_{{\bf r}'}G(E,{\bf P};{\bf r},{\bf r}') g({\bf r}') \psi({\bf r}'), 
\end{equation}
where the Green's function $G$ is defined by
\begin{equation}
\label{Green's defined}
G(E,{\bf P};{\bf r},{\bf r}') = \frac{1}{{\cal N}}\sum_{\bf q} 
\frac{\psi({\bf r}|{\bf q})\ \psi^*({\bf r}'|{\bf q})}
{E-E_{{\bf P},{\bf q}}}.
\end{equation}

In what follows, we shall truncate the interaction potential so that 
$g({\bf r})$
vanishes beyond a certain distance, that is, $g({\bf r})\neq 0$ only when 
${\bf r}=0,\pm{\bf R}_1,
\ldots,\pm{\bf R}_{n-1}$. Equation (\ref{psi = G g psi}) for these sites can 
then be cast into matrix form:
\begin{equation}
\label{matrix form}
\left[1 - {\cal G}(E,{\bf P})g\right]\psi = 0, 
\end{equation}
where $\psi$ is an $n\!\times\! 1$ matrix with $\psi_j = \psi({\bf R}_j)$, 
${\cal G}(E,{\bf P})$ is an $n\!\times\! n$ matrix with 
${\cal G}_{ij}(E,{\bf P}) 
= G(E,{\bf P};{\bf R}_i,{\bf R}_j)$, and $g$ is a diagonal $n\!\times\! n$ 
matrix 
with $g_{ii}=g({\bf R}_i)$. Note that the kinetic energy of two particles 
(\ref{E(p)}) can also be written in matrix form, 
\begin{eqnarray}
E_{{\bf P},{\bf p}}&=&\epsilon_{({\bf P}/2)+{\bf p}} 
+ \epsilon_{({\bf P}/2)-{\bf p}} 
\nonumber \\
\label{KE matrix}
&=& 2\sum_{i,j}\psi^*_i({\bf P}/2)\tau_{ij}\psi_j({\bf p}),
\end{eqnarray}
where the diagonal matrix $\tau$ is defined by $\tau_{ii} = t({\bf R}_i)$ and 
where $\psi_i({\bf p})\equiv\psi({\bf R}_i|{\bf p})$.
It follows from Equation (\ref{matrix form}) that the
eigenvalues $E$ are determined by the secular equation, 
\begin{equation}
\label{secular}
\det{\left[{\cal G}(E,{\bf P}) - g^{-1}\right]} = 0.
\end{equation}
For each eigenvalue $E$, the wave function at the $n$ sites 
$\psi({\bf R}_j)$
is determined by Equation (\ref{matrix form}), while for sites beyond the 
interaction range one uses Equation (\ref{psi = G g psi}) to determine 
$\psi({\bf r})$.

In the case of a scattering process, the solution to the Schr\"odinger 
equation (\ref{Schrodinger}) with an incident plane wave 
$\psi({\bf r}|{\bf p})$ satisfies 
\begin{equation}
\label{Lippmann Schwinger}
\psi({\bf r}) = \psi({\bf r}|{\bf p}) + 
\sum_{{\bf r}'} G(E,{\bf P};{\bf r},{\bf r}')g({\bf r}')\psi({\bf r}').
\end{equation}
$\psi({\bf r})$ is therefore given by
\begin{eqnarray}
\label{scattered state}
\psi({\bf r}) &=& \psi({\bf r}|{\bf p}) \\
&+& \sum_{i=0}^{n}\sum_{j=0}^{n}G(E,{\bf P};{\bf r},{\bf R}_i)
{\cal A}_{ij}(E,{\bf P})\psi({\bf R}_j|{\bf p}), \nonumber
\end{eqnarray}
where the $n\!\times\! n$ scattering matrix ${\cal A}(E,{\bf P})$ is 
{\em defined} by the equation\footnote{It should be noted 
that ${\cal A}_{ij}\neq A({\bf r},{\bf r}')|_{{\bf r}
={\bf R}_i,{\bf r}'={\bf R}_j}$, where $A$ is
the scattering matrix defined on the entire lattice.  It is true, however, 
that ${{\cal A}^{-1}}_{ij} = 
A^{-1}({\bf r},{\bf r}')|_{{\bf r}={\bf R}_i,{\bf r}'={\bf R}_j}$.} 
\begin{equation}
\label{scattering matrix}
{\cal A}^{-1}(E,{\bf P}) = g^{-1}-{\cal G}(E,{\bf P}).
\end{equation}
${\cal A}(E,{\bf P})$, as a function of complex $E$, has a cut along the real 
axis corresponding to the continuum spectrum of two free fermions. 
Depending on the potential $g({\bf r})$, it may also have poles on the real 
axis (bound states) or below the upper edge of the cut (resonances). 

\subsection{Symmetry of the ground state}
\label{symmetry of the ground state}

We shall now demonstrate that 
the wave function of the ground state of two fermions does not change 
sign unless some of the hopping amplitudes $t({\bf r})$ are {\em positive}.
The situation 
is similar to that in the continuum, for which the ground-state wave 
function has no nodes. Note that a positive effective mass [Equation 
(\ref{fermion mass})] requires at least some of the hopping amplitudes to 
be negative.

We have seen that the problem of two fermions reduces to a 
one-body problem with modified hopping amplitudes.
It therefore suffices to study the ground state of 
a single particle moving on the same lattice in the external potential 
$g({\bf r})$. We consider the case of total momentum ${\bf P}=0$, 
for which the 
amplitudes double (reminiscent of the reduced mass effect in the continuum). 
The ground state is obtained by minimizing the expectation value of 
the energy:
\begin{equation}
\label{minimization}
\langle E\rangle 
= \sum_{{\bf r},{\bf r}'}2t({\bf r}-{\bf r}')\psi^*({\bf r}')\psi({\bf r}) 
+ \sum_{\bf r} g({\bf r})|\psi({\bf r})|^2.
\end{equation}

Suppose that $t({\bf r})\leq 0$ for all ${\bf r}$ and that $\psi({\bf r})$ 
is positive 
on some site ${\bf r}_1$ and negative on another site ${\bf r}_2$. 
Then the first 
term in (\ref{minimization}) can 
be lowered if instead of a negative $\psi({\bf r}_2)$ we use 
$|\psi({\bf r}_2)|$, 
the second term remaining unaffected by such a change.
Therefore, the absolute minimum of $\langle E\rangle$ must have 
$\psi({\bf r})\geq 0$ on all sites, provided there always exists a sequence of 
allowed hops from any one site to any other site on the lattice. 

\section{Deep bound states}\label{deep}

When the magnitude of an attractive potential exceeds the 
kinetic energy of two fermions, 
compact pairs are formed and can be treated as small-size 
bosons.  Pairs formed by two fermions on the same site, however, 
have a vanishing mobility in the limit of strong binding. This 
strongly decreases the condensation temperature \cite{Nozieres},  
in contrast to the continuum case.  It is desirable to have a 
model which does not have this lattice artifact. The pairing of 
fermions on adjacent sites is one way around the difficulty.
We address this issue by working out two specific examples of lattice 
fermions having a strong nearest-neighbor attraction: 
first on a face-centered cubic lattice and then on a square lattice.

The advantage of extended-boson models is offset by a new 
problem, which is also a lattice effect.  While only two fermions
can be involved in on-site attractive interactions due to the 
Pauli principle, extended-range attractive potentials can result in 
the formation of many-body bound states. However, quantum statistics can 
help avoid this type of phase separation because the interaction strength 
in symmetric (spin-singlet) and antisymmetric (spin-triplet) spatial states 
can be quite different.  Let us assume that the interaction 
is attractive in a singlet state and repulsive in a triplet.  
Because a fermion 
cannot form more than one singlet bond at a time, it will attract
only one other fermion.  More exactly, quantum spin fluctuations 
will produce a fluctuating potential between fermions.  Only 
in pairs are these fluctuations completely suppressed, and 
a state with pairs may be favored over other many-body states.  For details,
see Appendix \ref{many-body}.

\subsection{Deep bound states on a face-centered cubic lattice}
\label{fcc}

The following simple model describes tightly bound fermion pairs 
with a small effective mass (close to that of individual fermions).
The small mass results from fermionic pairs which move through the lattice 
without having to break the bond. Triangular and f.c.c. lattices 
offer such a possibility. 

The Hamiltonian [Equations (\ref{kinetic energy}) and (\ref{interaction})] 
for the f.c.c. model contains nearest-neighbor hopping with amplitude $t$, 
on-site repulsion $g_0$, and attraction $g_1$ for fermions on adjacent sites.
On an f.c.c. lattice, there are twelve nearest neighbors ${\bf r}'$ 
around any one site ${\bf r}$, with 
\begin{equation}
\label{fcc neighbors}
{\bf r}'-{\bf r} = 
\frac{a}{2}(\pm\hat{\bf x}\pm\hat{\bf y}),\ 
\frac{a}{2}(\pm\hat{\bf y}\pm\hat{\bf z}),\
\frac{a}{2}(\pm\hat{\bf z}\pm\hat{\bf x}).
\end{equation}
$\hat{\bf x},\ \hat{\bf y}$, and $\hat{\bf z}$ are the unit vectors 
along each side of the cube and $a$ is the length of each side.

If $|g_1|\!\gg\!|t|$, so that the bond between fermions is strong, 
the relative-position wave function $\psi({\bf r})\neq0$ only on the 12 sites 
surrounding the origin. We therefore expect 12 bound states with 
the same potential energy $g_1$ and, in general, unequal kinetic 
energies 
at a total Bloch momentum ${\bf P}\neq0$. The kinetic energy is
determined by the 
hopping amplitudes, $2t\cos{({\bf P}\!\cdot\!{\bf r}/2)}$, as discussed 
in Section \ref{basic notations}. 

\begin{figure}
\begin{center}
\epsfxsize=3.125in
\leavevmode
\epsffile{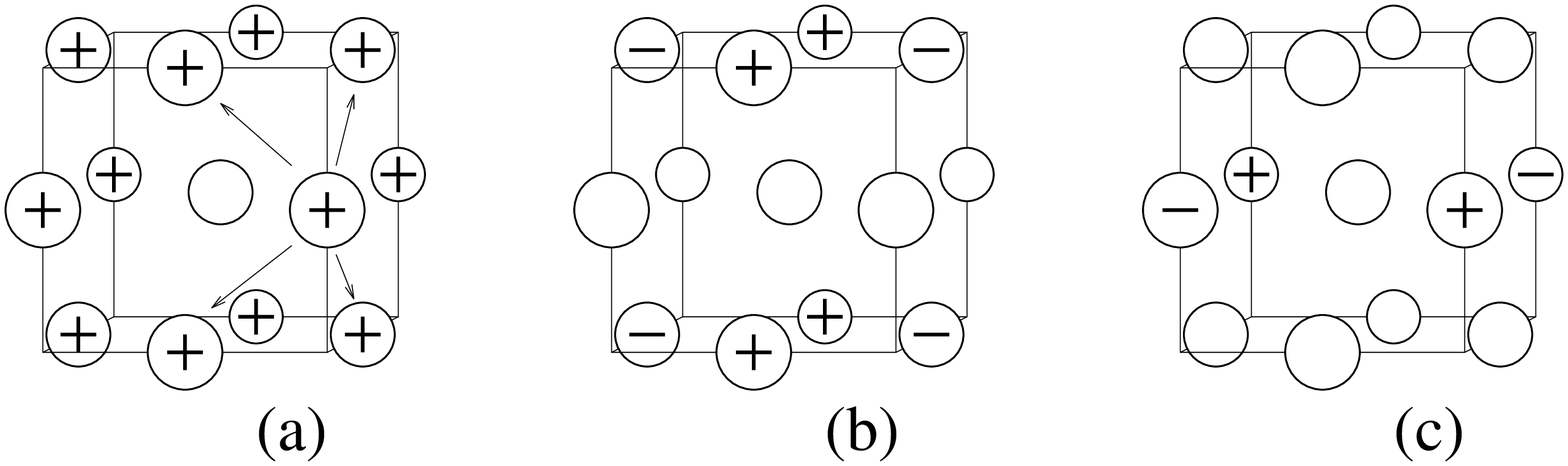}
\end{center}
\caption{Deep bound states on an f.c.c. lattice.
Symmetries and energies of the states:  (a) $A_1$, 
$g_1+8t$. (b) $E$, $g_1-4t$. (c) $T_2$, $g_1$.}
\label{fcc states}
\end{figure}

At ${\bf P}=0$, all hopping amplitudes are $2t$ and the energy 
levels have degeneracy (see Appendix \ref{symmetry appendix}):
there are a singlet $A_1$, a doublet $E$, and a triplet $T_2$ 
(Figure \ref{fcc states}). As long as $t<0$, the ground state for 
${\bf P}=0$ is $A_1$,  
with $\psi({\bf r})=1/\sqrt{12}$ on the 12 sites around the 
origin and zero otherwise. From each of these 12 sites a fermion can 
hop to 4 of its nearest neighbors (see Figure \ref{fcc states}(a)). 
For ${\bf P}\neq0$, the boson dispersion relation to lowest order in 
${\bf P}$ is given by expanding $\cos{({\bf P}\!\cdot\!{\bf r}/2)}$ to 
second order:
\begin{equation}
E({\bf P}) = g_1 + 8t -\frac{t}{12}\sum_{{\bf r},{\bf r}''}^{\rm n.n.}
\left[\frac{{\bf P}\!\cdot\!({\bf r}-{\bf r}'')}{2}\right]^2,
\end{equation}
where ${\bf r}''$ denotes the four nearest-neighbor sites shown in Figure 
\ref{fcc states}(a).
Because $E$ is an isotropic function of ${\bf P}$ to this order, we average 
over
all directions of the total momentum and substitute 
$|{\bf r}-{\bf r}''|=a/\sqrt{2}$ 
to obtain
\begin{equation}
E({\bf P}) = g_1 + 8t -\frac{tP^2a^2}{6},
\end{equation}
which gives the effective mass of a boson as 
\begin{equation}
\label{fcc mass}
m_b = -\frac{3\hbar^2}{t a^2} = 6m_f.
\end{equation}

\subsection{Deep bound states on a square lattice.}
\label{deep square}

On a square lattice, the mass of a strongly bound pair will be finite 
if one takes into account next-nearest-neighbor fermion hopping (with
amplitude $t_2$) in addition to nearest-neighbor hopping (with amplitude 
$t_1$). Suppose the interaction potential has 
strength $g_0>0$ for particles on the same site, $g_1<0$ for nearest 
neighbors, $g_2>0$ for next-nearest neighbors, and 0 otherwise. For spin 
singlets, we have five distinct relative positions with non-zero interactions:
${\bf R}_0=(0,0),\ {\bf R}_1=(0,\pm a),\ {\bf R}_2=(\pm a, 0),\ 
{\bf R}_3=(\pm a, \pm a)$, and 
${\bf R}_4=(\pm a, \mp a)$. In this basis, 
$\tau = {\rm diag}(0,t_1,t_1,t_2,t_2)$ 
and $V = {\rm diag}(g_0,g_1,g_1,g_2,g_2)$. According to 
Equation (\ref{fermion mass}), the fermion mass is 
\begin{equation}
m_f = \frac{\hbar^2}{2(-t_1-2t_2)a^2}.
\end{equation}

As in the determination of a single fermion mass, we expand the
energy of a two-fermion bound state in powers of ${\bf P}$.
Near ${\bf P}=0$, it is convenient to use the irreducible representations of 
the symmetry group (see Appendix \ref{symmetry appendix}); and we 
choose the following five 
components $\psi_i({\bf p})$, 
\begin{equation}
\label{5 components}
\psi({\bf p}) \equiv \left(\begin{array}{c} 
\psi_0({\bf p})\\\psi_1({\bf p})\\
\psi_2({\bf p})\\\psi_3({\bf p})\\
\psi_4({\bf p})\end{array}\right) = 
\left(
\begin{array}{c}
1 \\ \cos{p_xa} + \cos{p_ya} \\ \cos{p_xa} - \cos{p_ya} \\ 
2\cos{p_xa}\ \cos{p_ya} \\ 2\sin{p_xa}\ \sin{p_ya}
\end{array} 
\right), 
\end{equation}
as a basis set. Functions  
$\psi_0$, $\psi_1$, and $\psi_3$ belong to the $A_1$ representation 
(refered to as $s$-wave in the literature); 
$\psi_2$ transforms as $B_1$ ($d_{x^2-y^2}$-wave); and $\psi_4$ as $B_2$ 
($d_{xy}$-wave). Assuming that the 
nearest-neighbor attraction $g_1$ is strong and $E\approx g_1$, we expand 
the Green's function
\begin{equation}
G_{ij}(E,{\bf P})\equiv\frac{1}{\cal N}
\sum_{{\bf p}}\frac{\psi_i({\bf p}) \psi^*_j({\bf p})}
{E-2\sum_{k,l=0}^{4}\psi^*_k({\bf P}/2)\tau_{kl} \psi_l({\bf p})}
\end{equation}
in powers of $1/E$ and thus determine the dependence $E({\bf P})$. Because
\begin{equation}
G_{ij}(E,{\bf P}) = \frac{\delta_{ij}}{E} 
+ {\cal O}\left(\frac{1}{E^2}\right),
\end{equation}
we need only the diagonal matrix elements to calculate the lowest-order 
correction to $E$. We have 
\[G_{00}(E,{\bf P})=G_{33}(E,{\bf P})=G_{44}(E,{\bf P})
=\frac{1}{E}+{\cal O}\left(\frac{1}{E^3}\right), \]
\[G_{11}(E,{\bf P}) 
=\frac{1}{E} + \frac{2t_2}{E^2}\psi_3({\bf P}/2) 
+{\cal O}\left(\frac{1}{E^3}\right),\] 
\[ G_{22}(E,{\bf P}) 
=\frac{1}{E} - \frac{2t_2}{E^2}\psi_3({\bf P}/2) 
+{\cal O}\left(\frac{1}{E^3}\right).\]
Upon substituting these expressions into equation (\ref{secular}), we obtain 
the following dispersion laws:
\begin{equation}
\label{dispersion}
E = g_1 \pm2t_2\psi_3({\bf P}/2) + {\cal O}\left(\frac{1}{g_1}\right),
\end{equation}
where the plus sign corresponds to the $A_1$ state $\psi_1$ and the minus 
sign corresponds
to the $B_1$ state $\psi_2$. The other three bands have no dispersion to this 
order.
For a positive (negative)
$t_2$, the ground state transforms according to representation $B_1$ ($A_1$) 
and the mass of the boson is determined strictly by the diagonal hopping term:
\begin{equation}
m_b = \frac{\hbar^2}{|t_2|a^2} = \frac{-2t_1-4t_2}{|t_2|}m_f.
\end{equation}
Note that, unlike in the case of a strong {\em on-site} attraction, an
extended boson can easily move through the lattice without breaking its
bond, hence a small mass even for a strong attraction.  It can even be 
smaller than the mass of its constituent fermions. 

It is straightforward to see what determines the symmetry of the ground 
state. 
The relative position of two fermions is restricted to four nearest-neighbor
sites by a strong attraction. Depending on the sign of $t_2$, the wavefunction
must either alternate its sign on these sites or maintain its sign to 
minimize the kinetic energy.

\section{Shallow bound states}
\label{shallow}

In this section, we investigate the properties of bound states that lie 
close to the bottom of the two-fermion band, which is assumed to 
have a single minimum $E_0$ at the relative momentum ${\bf p}=0$. 
As described in Section \ref{Schrodinger equation}, we first solve Equation 
(\ref{matrix form}) on the relative sites having a non-zero interaction 
potential.
As before, we assume that $g({\bf r})\neq 0$ only for a finite number 
of lattice 
sites ${\bf r} = 0, \pm{\bf R}_1, \ldots, \pm{\bf R}_{n-1}$. 
Let us also assume that 
the same-site repulsion is infinite so that $\psi(0)=0$. 
Using Greek indices 
to enumerate sites $\pm{\bf R}_1, \ldots, \pm{\bf R}_{n-1}$, 
we obtain from Equation 
(\ref{matrix form}) the following set of equations for the vector 
$\Delta = g\psi:$\footnote{The BCS gap $\Delta({\bf r})$ is the 
second-quantized counterpart of this function.}
\begin{mathletters}
\begin{eqnarray}
{\cal G}_{00}\Delta_0 + {\cal G}_{0\alpha}\Delta_\alpha 
 &=& {g^{-1}}_{00}\Delta_0 = 0, \\ 
{\cal G}_{\alpha0}\Delta_0 + {\cal G}_{\alpha\beta}\Delta_\beta 
 &=& {g^{-1}}_{\alpha\beta}\Delta_\beta.
\end{eqnarray}
\end{mathletters}
Summation over repeated indices is assumed. Eliminating $\Delta_0$, we 
obtain 
\begin{equation}
\label{0 eliminated}
\left({\cal G}_{\alpha\beta}-\frac{{\cal G}_{\alpha0}{\cal G}_{0\beta}}
{{\cal G}_{00}}\right)\Delta_\beta 
= {g^{-1}}_{\alpha\beta}\Delta_\beta.
\end{equation}
Now we can use Equation (\ref{psi = G g psi}) to calculate 
the wave function on the remainder of the lattice: 
\begin{eqnarray}
\psi({\bf r}) &=& G({\bf r},{\bf R}_i)\Delta_i\nonumber\\
\label{the rest of the lattice}
&=&\left[G({\bf r},{\bf R}_\alpha) - \frac{G({\bf r},0)G(0,{\bf R}_\alpha)}
{G(0,0)}\right]\Delta_\alpha.
\end{eqnarray}
Apart from a multiplicative constant, the long-range behavior of 
$\psi({\bf r})$ 
is determined by the properties of the Green's function at large separations, 
which in turn is controlled by the long-wavelength behavior of the summand 
in Equation (\ref{Green's defined}). The latter diverges at 
${\bf q}=0$ as energy $E$ 
approaches the bottom of the two-fermion band. Then the denominator 
in (\ref{Green's defined}) can be approximated using the 
effective fermion mass:
\begin{equation}
\label{xi0 defined}
E - E_{{\bf P},{\bf q}} \approx E - E_{{\bf P},0} - \frac{q^2}{m_f} 
 = - \frac{q^2 + 1/\xi_0^2}{m_f}, 
\end{equation}
which defines a coherence length $\xi_0$ at themperature 
$T=0$ as the size of the bound 
state. For $r\gg\xi_0$, the propagator decays exponentially with the distance,
while in the intermediate range, $a\ll r\ll\xi_0$, it decreases as a power of 
$|{\bf r}-{\bf r}'|$.

It is possible to isolate the divergent part of the Green's 
function in the form of $G(E,{\bf P};0,0)\equiv G(0,0)$. We define 
a regularized Green's function as follows.
For ${\bf r}\neq0$ and ${\bf r}'\neq0$,
\begin{mathletters}
\begin{equation}
F({\bf r},{\bf r}') = \frac{1}{{\cal N}}\sum_{\bf q} \frac
{[\psi({\bf r}|{\bf q})-\psi({\bf r}|0)][\psi^*({\bf r}'|{\bf q})
-\psi^*({\bf r}'|0)]}{E-E_{{\bf P},{\bf q}}},
\end{equation}
\begin{equation}
F({\bf r},0) = \frac{1}{{\cal N}}\sum_{\bf q} \frac
{[\psi({\bf r}|{\bf q})-\psi({\bf r}|0)]\psi^*(0|{\bf q})}{E-E_{{\bf P},
{\bf q}}}, 
\end{equation}
\begin{equation}
F(0,{\bf r}') = \frac{1}{{\cal N}}\sum_{\bf q} \frac
{\psi(0|{\bf q})[\psi({\bf r}'|{\bf q})-\psi^*({\bf r}'|0)]}{E-E_{{\bf P},
{\bf q}}}.
\end{equation}
\end{mathletters}
These sums are regular at ${\bf q}=0$. By inspection, 
\begin{mathletters}
\begin{eqnarray}
G({\bf r},0) &=& \psi({\bf r}|0)G(0,0) + F({\bf r},0), \\ 
G(0,{\bf r}') &=& G(0,0)\psi^*({\bf r}'|0) + F(0,{\bf r}'), \\ 
G({\bf r},{\bf r}') &=& \psi({\bf r}|0)G(0,0)\psi^*({\bf r}'|0) 
+ \psi({\bf r}|0)F(0,{\bf r}')
\nonumber\\
&&{} + F({\bf r},0)\psi^*({\bf r}'|0) + F({\bf r},{\bf r}').
\end{eqnarray}
Upon substituting these functions into Equation 
(\ref{the rest of the lattice}), 
we finally obtain 
the wave function as a sum of two terms,  
\end{mathletters}
\begin{eqnarray}
\psi({\bf r}) &=& F(E,{\bf P};{\bf r},{\bf R}_\alpha)\Delta_\alpha \nonumber\\
\label{psi}
&&- F(E,{\bf P};0,{\bf R}_\alpha)\Delta_\alpha 
\frac{G(E,{\bf P};{\bf r},0)}{G(E,{\bf P};0,0)}.
\end{eqnarray}
Consider states with total Bloch momentum ${\bf P}\!=\!0$.
The first term in Equation (\ref{psi}) is proportional to the regularized 
propagator 
$F(E,{\bf P};{\bf r},{\bf R}_\alpha)$, 
whose Fourier transform remains finite as ${\bf q}\rightarrow0$. 
In the intermediate
range, $a\ll r\ll\xi_0$, it decays with ${\bf r}$ as $r^{-D}$ 
or faster ($D$ is 
the number of spatial dimensions). The second term, on the other hand, 
is on the order of 
$r^{2-D}$, as can be inferred from its singular Fourier transform. It therefore
should dominate at large distances. However, this term vanishes if 
the state 
belongs to any irreducible representation other than the trivial one due to 
the prefactor $F(E,0;0,{\bf R}_\alpha)\Delta_\alpha$.\footnote{The 
Green's function $F(E,0;0,{\bf R}_\alpha)$ is 
invariant under all point group transformations.} This means that 
a bound state which transforms according to a non-trivial irreducible 
representation of the point group will be smaller in size than  
a symmetric bound state of the same energy. This may have verifiable 
consequences for superconductivity. 

To illustrate the importance of this point, we use a continuum 
approximation to the two-particle problem. 
In the intermediate region, $a\!\ll\! r\!\ll\! \xi_0$, 
the two-particle wavefunction 
$\psi({\bf r})$ satisfies the continuum Laplace equation, 
$\nabla^2\psi({\bf r}) = 0$.
In 2 and 3 dimensions, solutions with angular momentum $l$ and its 
projection $m$ are
\begin{mathletters}
\begin{eqnarray}
\psi({\bf r}) &=& e^{im\phi}r^{-m} \mbox{ (2 dimensions)},\\
\psi({\bf r}) &=& Y_{lm}(\theta,\phi)r^{-l-1} \mbox{ (3 dimensions)}. 
\end{eqnarray}
It is generally thought that the lattice spacing $a$ is 
irrelevant as long as we deal with a weakly bound pair ($\xi_0\gg a$). 
A simple scaling 
argument demonstrates that this is not always the case. To imitate the 
infrared and ultraviolet cutoffs, we set $\psi({\bf r})=0$ outside the 
interval 
$a<r<\xi_0$. In $3$ dimensions, the $n$-th moment of $r$ is then given by 
\end{mathletters}
\begin{equation}
\label{nth moment}
\langle r^n\rangle = \frac{\int_{a}^{\xi_0}r^{-2l-2+n}r^2dr}
{\int_{a}^{\xi_0}r^{-2l-2}r^2dr}.
\end{equation}
For $l=0$ one obtains 
\begin{equation}
\label{l=0}
\langle r^n\rangle = \frac{\xi_0^n}{n+1}
\end{equation}
for any $n\ge0$. Indeed, an $s$ state is not sensitive to the ultraviolet 
cutoff (the lattice spacing). For $l\ge1$, 
\begin{mathletters}
\begin{eqnarray}
\label{n<2l-1}
\mbox{$n<2l-1$: }\ 
\langle r^n\rangle &=& \frac{2l-1}{2l-1-n}\ a^n,\\
\label{n=2l-1}
\mbox{$n=2l-1$: }\ 
\langle r^{n}\rangle &=& (2l-1)\ a^{2l-1}\log{\frac{\xi_0}{a}},\\
\label{n>2l-1}
\mbox{$n>2l-1$: }\ 
\langle r^n\rangle &=& 
\frac{2l-1}{n-2l+1}\ \xi_0^n\!\left(\frac{a}{\xi_0}\right)^{2l-1}.
\end{eqnarray}
This suggests that, for a shallow bound state with non-zero angular momentum,
the measured size of a pair is determined by both length scales, $\xi_0$ 
and $a$, and therefore may be much smaller than $\xi_0$. The physical reason 
behind this difference is that, at the same value of kinetic energy,
states with a larger {\em angular} momentum have a smaller {\em radial} 
momentum.  Therefore, the ability to tunnel into the classically forbidden 
region (to distances on the order of $\xi_0$) is substantially reduced.
Similar arguments hold for the two-dimensional case.
\end{mathletters}

The small size of bosonic pairs may imply a high upper 
critical magnetic field. As has been argued in \cite{FLR91}, 
the upper critical 
magnetic field for the condensate of an ideal charged Bose gas at 
zero temperature is infinite, because of the very different roles 
played by the Coulomb interaction in Bose and Fermi systems. 
Therefore, a rapid increase of the critical field
as $T\to 0$ should be expected for an ideal charged Bose gas. Recently, a 
similar behavior has been observed in $T_c$-suppressed cuprates
\cite{H_{c2}}. This 
observation indicates that Bose condensation may be dominant in 
these superconductors.  Of course, bosons in real systems are not 
point particles.  A rough estimate of the upper critical 
field would be the magnetic field at which the cyclotron radius of 
fermions is comparable to the size of d-wave pairs, 
\begin{equation}
\phi_0 = 2\pi H_{\rm c2} \langle r^2\rangle,
\end{equation}
where $\phi_0=hc/2e$ is a flux quantum.  Formally, this criterion reproduces
the Landau--Ginzburg result, provided that the pair size is replaced
by the coherence length $\xi$. We stress, however, that in a $d$-wave 
superconductor the former can be smaller than the latter.


We now investigate the model described in Section \ref{deep square} 
in order to determine the symmetry of a shallow ground state.  Consider 
the case in which 
the bound state first appears below the continuum band. 
Using the basis set (\ref{5 components}), we write the secular equation 
in the form of Equation (\ref{0 eliminated}).  It is assumed that the on-site
repulsion $g_0$ is infinite.  For a state with an energy just at 
the bottom of the two-fermion band, it is convenient to use 
a regularized Green's function ${\cal F}(E,{\bf P})$ as was introduced previously.  
Equation (\ref{0 eliminated}) then reads 
\begin{equation}
\label{0 eliminated via F}
\left({\cal F}_{\alpha\beta}-\frac{{\cal F}_{\alpha0}{\cal F}_{0\beta}}
{{\cal G}_{00}}\right)\Delta_\beta 
= {g^{-1}}_{\alpha\beta}\Delta_\beta.
\end{equation}
${\cal G}_{00}$ becomes infinite as $E\to 2\epsilon_0$, which further 
simplifies the problem.  The critical values of the coupling strength
are then given by the equation
\begin{equation}
\label{critical g}
\left({\cal F}_{\alpha\beta}-{g^{-1}}_{\alpha\beta}\right)
\Delta_\beta = 0,
\end{equation}
where ${\cal F}_{\alpha\beta}$ is to be evaluated at ${\bf P}=0$ 
and $E=2\epsilon_0$.  For the irreducible representations $A_1$, $B_1$, and 
$B_2$; one obtains, respectively,
\begin{mathletters}
\begin{eqnarray}
\label{s}
({\cal F}_{11}-g_1^{-1})({\cal F}_{33}-g_2^{-1}) 
&=& {\cal F}_{13}{\cal F}_{31},\\
\label{d_x^2-y^2}
{\cal F}_{22} - g_1^{-1} &=& 0,\\
\label{d_xy}
{\cal F}_{44} - g_2^{-1} &=& 0.
\end{eqnarray}
\end{mathletters} 
As long as $g_2$ represents repulsion, the $B_2$ bound state 
($d_{xy}$) does not exist.  

If there is no next-nearest-neighbor repulsion ($g_2=0$), it can be 
seen that the ground state is $A_1$ ($s$ wave).  Indeed, Equation (\ref{s}) 
reduces to ${\cal F}_{11}=g_1^{-1}$ and one obtains the following relation
between the critical values of $g_1$ for $s$ and $d_{x^2-y^2}$ waves:
\begin{equation}
\frac{1}{g_1^{(s)}} - \frac{1}{g_1^{(d)}} = 
\frac{2}{{\cal N}}\sum_{\bf q}\frac{(1-\cos{q_x a})(1-\cos{q_y a})}
{\epsilon_0-\epsilon_{\bf q}} < 0.
\end{equation}
Thus, it takes a weaker attraction to produce an $s$-wave ground state
when there is no next-nearest-neighbor repulsion. 

\begin{figure}
\begin{center}
\input{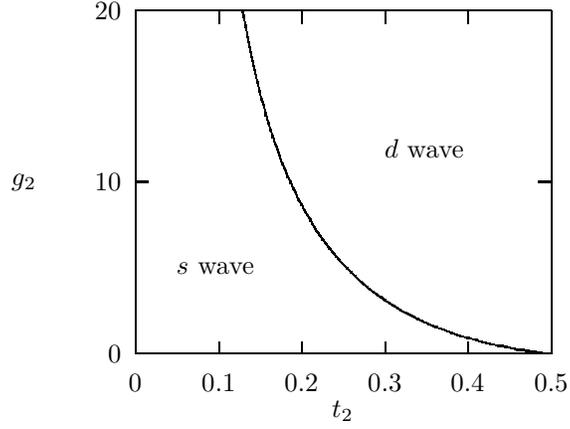}
\end{center}
\caption{Shallow $s$-wave and $d$-wave ground states on a square lattice. 
Nearest-neighbor hopping $t_1=-1$.}  
\label{s versus d}
\end{figure}

When the next-nearest-neighbor amplitude $t_2$ is negative, the ground 
state is an $s$ wave, in accordance with the result of Section 
\ref{symmetry of the ground state}.  Therefore, the search for 
a non-trivial symmetry of the ground state is restricted to the 
domain in which both $t_2>0$ and $g_2>0$.  The results can be 
summarized as follows.  An increasing next-nearest-neighbor 
repulsion $g_2$ raises the kinetic energy of an $s$ wave but does 
not affect that of a $d_{x^2-y^2}$ wave.  A positive $t_2$ means 
positive kinetic energy for $s$ waves and negative kinetic energy for 
$d$ waves. 
The actual ``phase diagram'' is presented in Figure \ref{s versus d}.
For the value of $t_2=0.45|t_1|$ inferred by Pines from 
fitting NMR results \cite{Pines}, we find that the 
strength of the next-nearest-neighbor repulsion must be at least 
$0.35|t_1|$ in order for the ground state to have the 
$(x^2-y^2)$ symmetry.  

\section{Extended resonance states}\label{resonance}

\subsection{General approach}

Even if the two-particle potential energy is positive at all separations,
there still remains the possibility of two fermions 
forming a resonance state -- a state bound for a limited time -- provided that 
the potential has a local minimum. It has been shown \cite{FL89} 
that resonance bosons can form a condensate, thus inducing 
superconductivity in the boson--fermion system.

On a lattice, we may approach the problem of a single resonance in the 
following way. When a local minimum of the potential is surrounded by 
a closed wall of sites with infinite repulsion, the wave function cannot leak
out and two fermions form a stable bound state. Then
the Schr\"odinger equation, $(H_0+H_1)|{\bf P},\psi\rangle 
= E|{\bf P},\psi\rangle$,
has a solution for a real $E$ inside the two-fermion energy band. The 
wave function $\psi({\bf r})$ is non-zero only for separations ${\bf r}$
within the wall, which makes the Schr\"odinger equation readily solvable. 
Its solution also satisfies Equation 
(\ref{psi = G g psi}), which determines the location of the pole of the 
scattering amplitude. In the limit of a stable state, the residue of the pole 
vanishes, reflecting the fact that incident free particles cannot penetrate 
the barrier of infinite repulsion. 
Lowering the height of the wall to a finite value makes the localized 
two-fermion state unstable, thereby shifting the pole away from the real axis 
into the unphysical sheet. 
In what follows, we determine the mass and resonance width of a bosonic state 
to lowest orders in the inverse height of the wall. 

Let the wave function of the stable state be $\psi({\bf r})$, normalized to 1. 
Outside the wall, $\psi({\bf r})=0$. $\psi({\bf r})$ satisfies 
Equation (\ref{psi = G g psi}),
which now must be modified due to the existence of a branch cut of the 
function $G(E,{\bf P};{\bf r}',{\bf r})$ on the real axis of $E$: 
\begin{mathletters}
\begin{equation}
\label{psi = G g psi near cut}
\psi({\bf r}) = 
\sum_{{\bf r}'}G(E\pm i0^+,{\bf P};{\bf r},{\bf r}') g({\bf r}') 
\psi({\bf r}')
\end{equation}
Because $\psi({\bf r})$ does not contain incoming or outgoing free particles, 
it is invariant under time reversal and therefore satisfies this equation 
on both edges of the cut. The complex conjugate of 
(\ref{psi = G g psi near cut}) is 
\begin{equation}
\label{psi* = psi* g G near cut}
\psi^*({\bf r}) = \sum_{{\bf r}'}\psi^*({\bf r}')g({\bf r}')
G(E\mp i0^+,{\bf P};{\bf r}',{\bf r}).
\end{equation}
\end{mathletters}

As before, let us assume that the inter-particle potential  
$g({\bf r})\neq0$ only 
at the $2n-1$ sites ${\bf r}=0,\ \pm{\bf R}_1,\ldots,\pm{\bf R}_{n-1}$.  
We introduce a new 
wave function $\Delta({\bf r})\equiv g({\bf r})\psi({\bf r})$, 
which remains finite at a site 
${\bf r}$ even as $g({\bf r})\to\infty$ as long as the state has finite 
energy. Using the matrix notations of 
Section \ref{Schrodinger equation}, we rewrite Equations 
\ref{psi = G g psi near cut} 
and \ref{psi* = psi* g G near cut} in terms of the 
$n\!\times\! 1$ column vector $\Delta$: 
\begin{mathletters}
\begin{eqnarray}
\label{Schro0ket}
\left[g^{-1}-{\cal G}(E\pm i0^+,{\bf P})\right]\Delta &=& 0,\\
\label{Schro0bra}
\Delta^\dagger\left[g^{-1}-{\cal G}(E\mp i0^+,{\bf P})\right] &=& 0.
\end{eqnarray}
In what follows, we will omit the variables $E$ and ${\bf P}$ unless their 
presence is necessary to avoid ambiguity.
\end{mathletters}
 
By regarding the diagonal matrix $g^{-1}$ as an independent variable, we can 
treat the position of the pole $E$ as a function of $g^{-1}$. When 
$g^{-1}({\bf r})=0$ on the walls, $E$ is readily determined 
from the Schr\"odinger 
equation for the interior of the wall.   By changing $g^{-1}$ infinitesimally, 
we can trace the trajectory of $E$ in the complex plane. Differentiate 
Equation (\ref{Schro0ket}):
\begin{equation}
\label{Schro1ket}
(g^{-1}-{\cal G})d\Delta + 
\left(dg^{-1} - \frac{\partial{\cal G}}{\partial\!E}d\!E\right)\Delta
= 0.
\end{equation}
Multiplication of Equation (\ref{Schro1ket}) by $\Delta^\dagger$ eliminates 
the first term because of Equation (\ref{Schro0bra}). Then,
\begin{equation}
\label{dE raw}
\Delta^\dagger \frac{\partial {\cal G}}{\partial\! E}\Delta d\!E = 
\Delta^\dagger dg^{-1}\Delta.
\end{equation}
The factor in front of $d\!E$ is merely $-1$ (see Appendix 
\ref{resonance appendix}). 

We thus have the following expression for the first-order displacement of 
the pole:
\begin{equation}
\label{dE}
d\!E = -\Delta^\dagger dg^{-1}\Delta,
\end{equation}
which is real. Formally, $d\!E$ can be written as $\psi^\dagger dg\psi$, the
first-order term in the standard perturbation series, although
the latter is not well defined for an infinite repulsion.

Next, from Equation (\ref{Schro1ket}) we obtain the first-order correction to 
$\Delta$: 
\begin{eqnarray}
d\Delta &=& 
-{\cal A}\left(dg^{-1} - \frac{\partial {\cal G}}{\partial\! E}d\!E\right)
\Delta\nonumber\\ 
\label{dDelta}
&=& -{\cal A}\left(dg^{-1}g + {\cal G}\,d\!E\right)\psi,
\end{eqnarray}
where ${\cal A}^{-1} = g^{-1}-{\cal G}^{-1}$.
Here, again, the substitution of $\partial {\cal G}/\partial\! E$ by 
$-{\cal G}^2$ requires justification (Appendix \ref{resonance appendix}). 

To obtain the second-order correction to the energy of the pole $E$, we 
differentiate Equation (\ref{Schro1ket}) once again and multiply by 
$\Delta^\dagger/2$ to eliminate the term with $d^2\!\Delta$:
\begin{eqnarray}
&-&\frac{1}{2}\Delta^\dagger
\left(
\frac{\partial^2{\cal G}}{\partial\! E^2}d\!E^2 
+ \frac{\partial{\cal G}}{\partial\! E}d^2\!E 
\right)\Delta \nonumber\\
\label{Delta Schro2ket}\\
&+&\Delta^\dagger
\left( dg^{-1} - \frac{\partial{\cal G}}{\partial\! E}d\!E\right)
d\Delta = 0.\nonumber
\end{eqnarray}
The second-order differential of the independent variable $g^{-1}$ is zero. 
We simplify this expression by using 
$\Delta^\dagger(\partial{\cal G}/\partial\! E)\Delta=-1$ and 
\footnote{This relation can also be proved 
along the lines of Appendix \ref{resonance appendix}.  
For higher derivatives, however, one will have to use the values 
of the Green's function on the entire lattice. Therefore, this type of 
analysis only works through second order.}
$\Delta^\dagger(\partial^2{\cal G}/\partial\! E^2)\Delta
=2\psi^\dagger{\cal G}\psi$
and by substituting $d\Delta$ from (\ref{dDelta}): 
\begin{eqnarray}
\frac{1}{2}d^2\!E 
&=& \psi^\dagger{\cal G}\psi + 
\psi^\dagger(g\,dg^{-1} + {\cal G}\,d\!E)\,{\cal A}\,(dg^{-1}g 
+ {\cal G}\,d\!E)\psi 
\nonumber\\
\label{d2E ready}
&=& -\psi^\dagger g^{-1}\psi 
+ d\phi^\dagger{\cal A}\,d\phi, 
\end{eqnarray}
where 
\begin{equation}
\label{dphi}
d\phi \equiv (dg^{-1}g + g^{-1}d\!E)\psi 
= (dg^{-1} - \Delta^\dagger dg^{-1}\Delta g^{-2})\Delta.
\end{equation}
In Equation (\ref{d2E ready}) we used the definition 
${\cal A}^{-1}\!=\!g^{-1}\!-\!{\cal G}$ 
and the fact that 
$dg^{-1}\psi=0$ (the potential has been changed on infinite-repulsion sites  
where the wave function vanished). The first term in Equation 
(\ref{d2E ready}) is real. The imaginary part of the energy is thus given by 
the second term: 
\begin{equation}
\label{linewidth}
\frac{1}{2}{\rm Im} d^2\!E 
= {\rm Im} d\phi^\dagger {\cal A}\,(E\pm i0^+,{\bf P}) d\phi, 
\end{equation}

We now derive an expression for the mass of a resonance state. The problem of 
a resonance formed by two fermions on the same site has been addressed 
in an earlier work by Friedberg, Lee, and Ren \cite{FLR94}. 
They found that, as the 
lifetime of such a resonance increases, its mass grows without limit. 
This can be understood as follows. Two fermions constituting a 
long-lived resonance must be located on the same site. However, the 
motion of such a boson is mediated by  
consecutive hopping of the fermions to an adjacent site.  This means that the 
fermions must part and then reunite: the boson decays into two fermions 
and then reappears 
at an adjacent site. As the decay amplitude vanishes, so does the kinetic 
energy of the resonance, thus giving an infinite mass.  The same problem 
exists in the case of bound states \cite{Nozieres}.  
The situation changes if we consider a resonance formed by fermions 
on two adjacent sites. Now a pair of particles can move without changing the 
distance between them. On a triangular or face-centered cubic lattice, this 
can be accomplished with only nearest-neighbor hopping. On a square or 
simple cubic lattice, fermions must be able to hop to next-nearest-neighbor 
sites. Under such circumstances, even long-lived resonances will have 
a finite mass, which can readily be determined. In the limit of zero resonance 
width, the two fermions cannot penetrate the wall of infinite repulsion that 
holds them together. The wave function of such a state is non-zero 
for only a few relative separations. As was shown in Section 
\ref{basic notations}, the relative motion of two particles with total Bloch 
momentum ${\bf P}$ is equivalent to the motion of a single particle on the 
same lattice with hopping amplitudes 
$t'({\bf r}) = 2t({\bf r})\cos{({\bf P}\!\!\cdot\!{\bf r}/2)}$ 
in the external potential $g({\bf r})$.
As the total momentum departs from 0, the variation of the hopping terms  
is quadratic in ${\bf P}$. The second derivative of the energy with respect to 
${\bf P}$ gives the effective mass. Assuming that the wave function of the 
stable state $\psi({\bf r})$ is non-degenerate at ${\bf P}=0$, we treat the 
variation in the kinetic energy for small ${\bf P}$ as a perturbation.  
To lowest order, 
\begin{equation}
\label{KE, P>0}
E({\bf P}) = 
E(0) - \sum_{{\bf r},{\bf r}'}\psi^*({\bf r}')t({\bf r}-{\bf r}')\psi({\bf r})
\left[\frac{{\bf P}\!\!\cdot\!({\bf r}-{\bf r}')}{2}\right]^2,
\end{equation}
from which one obtains the mass. To this order, the mass is finite only 
for states with $\psi({\bf r})\neq 0$ on more than one site.  
When the state becomes unstable, its energy is altered. 
Since $\Delta({\bf r})$ is an implicit 
function of the total Bloch momentum (through the variation of the fermion 
kinetic energy), Equation (\ref{dE}) contains the next-order correction to the 
energy. This correction is proportional to the inverse height of the 
repulsion walls and is therefore larger than the linewidth (\ref{linewidth}).
Equation (\ref{dE}) determines the bosonic mass when (\ref{KE, P>0}) does not 
disperse with ${\bf P}$, as in the case of resonances formed 
on one single site. 

\subsection{Resonance on a square lattice: an example}

Here we illustrate the results from the previous section. 
Consider a resonance that is formed on a square lattice by two fermions on the 
same site. This is possible when the hopping range is restricted to 
nearest neighbors (amplitude $t$) and the inter-particle potential is large 
for two fermions on adjacent sites, $g_1\!\gg\!|t|$. The value of the 
potential $g_0$
on the same site controls the energy of the resonance and therefore 
should be within the continuum of two free fermions: 
$-8|t|<g_0<8|t|$. 

We start with a stable state, which is formed in the limit of $g_1\to\infty$. 
In this case, the bound state has the wave function 
$\psi({\bf r}) = \delta_{{\bf r}0}$ 
and the energy $E=g_0$. 
The value of $\Delta({\bf r})\equiv g({\bf r})\psi({\bf r})$ 
on sites with an infinite 
repulsion is then calculated by standard perturbation methods (with  
(\ref{KE eigen}) as a perturbation Hamiltonian):
\begin{equation}
\Delta({\bf r}) = \lim_{g_1\to\infty} g_1\psi({\bf r}) 
= - 2t\cos{({\bf P}\!\cdot\!{\bf r}/2)}. 
\end{equation}
We then have a $5\!\times\! 1$ column vector $\Delta$, 
\begin{eqnarray}
\Delta \equiv \left(\begin{array}{c}
\Delta(0,0)\\ \Delta(1,0)\\ \Delta(0,1)\\ \Delta(-1,0)\\ \Delta(0,-1)
\end{array}\right)
= \left(\begin{array}{c}
g_0\\-2t\cos{(P_xa/2)}\\-2t\cos{(P_ya/2)}\\-2t\cos{(P_xa/2)}\\
-2t\cos{(P_ya/2)} 
\end{array}\right).
\end{eqnarray}
For spin-singlet states, we only need that part which is symmetric under 
inversion of the relative separation ${\bf r}$. Performing a unitary 
transformation, we obtain a $3\!\times\! 1$ column vector for singlet states, 
\begin{equation} 
\Delta 
= \left(\begin{array}{r}
g_0\ u_0({\bf P}) \\ 
-2t\ u_1({\bf P})\\
-2t\ u_2({\bf P})
\end{array}\right),
\end{equation}
where 
\begin{equation}
\begin{array}{c}
u_0({\bf P}) = 1, \\ 
u_1({\bf P}) = \cos{(P_xa/2)}+\cos{(P_ya/2)},\\ 
u_2({\bf P}) = \cos{(P_xa/2)}-\cos{(P_ya/2)}.
\end{array}
\end{equation}
In the limit as $g_1\to\infty$, the matrix $g^{-1}$ is given by
\begin{equation}
g^{-1} = \left(\begin{array}{ccc} 
g_0^{-1}&0&0\\
0&0&0\\
0&0&0
\end{array}\right).
\end{equation}
A large but finite repulsion for nearest neighbors gives 
\begin{equation}
dg^{-1} = \left(\begin{array}{ccc}
0&0&0\\
0&g_1^{-1}&0\\
0&0&g_1^{-1}
\end{array}\right).
\end{equation}
We then obtain the following expression 
for the energy of a resonance through the first order in $t/g_1$:
\begin{eqnarray}
E({\bf P}) &=& E^{(0)} - \Delta^\dagger dg^{-1}\Delta \nonumber\\
\label{dispersion of resonance}
&=& g_0 - \frac{4t^2}{g_1}\left[u_1^2({\bf P}/2)+u_2^2({\bf P}/2)\right].
\end{eqnarray}
This gives the inverse effective mass of the resonance boson as
\begin{equation}
\frac{1}{m} = \frac{4t^2a^2}{g_1}
\end{equation}
and the width of the bosonic band as 
\begin{equation}
\label{bandwidth}
\Delta E = \frac{32t^2}{g_1}.
\end{equation}

To determine the resonance width, we use Equations.\ (\ref{linewidth}) and 
(\ref{dphi}) with
\begin{equation}
d\phi = -\frac{2t}{g_0g_1}\left(\begin{array}{c}
2t\left[u_1^2({\bf P}/2)+u_2^2({\bf P}/2)\right]\\
g_0 u_1({\bf P}/2)\\
g_0 u_2({\bf P}/2)
\end{array}\right).
\end{equation}
Evaluation of $d\phi^\dagger {\cal A} d\phi$ is complicated 
by the fact that its
inverse, ${\cal A}^{-1}\equiv g^{-1} -{\cal G}$, 
has a zero eigenvalue. Indeed, it 
follows from Equation (\ref{matrix form}) that 
\begin{equation}
\label{1/A Delta=0}
{\cal A}^{-1}\Delta = 0.
\end{equation}
However, by its definition, vector $d\phi$ is orthogonal to $\Delta$ and 
$d\phi^\dagger {\cal A} d\phi$ is not singular. 
The calculation simplifies considerably for zero Bloch momentum of the 
resonance: $u_2(0)=0$ and the linear space becomes two-dimensional.
In this case, $d\phi$ is also an eigenvector of the matrix 
${\cal A}^{-1}$, with  
eigenvalue $\lambda$.  The set of four equations, 
Equation (\ref{1/A Delta=0}) and 
${\cal A}^{-1}\ d\phi = \lambda\ d\phi$, contains three 
identities for the matrix 
${\cal G}$ (see Appendix \ref{Elliptic appendix}) and determines 
the eigenvalue 
$\lambda$. Solving for $\lambda$ yields 
\begin{equation}
\label{lambda}
\lambda = -\left(\frac{4t}{g_0}+\frac{g_0}{4t}\right){\cal G}_{10}.
\end{equation}
Upon substituting the expression (\ref{lambda}) 
into Equation (\ref{linewidth}) 
and expressing 
${\cal G}_{10}$ in terms of the simpler quantity 
${\cal G}_{00}$, we finally obtain 
\begin{equation}
\label{Gamma/2}
{\rm Im} E = \left(\frac{16t^2}{g_1}\right)^2
\frac{{\rm Im} {\cal G}_{00}}{|1-g{\cal G}_{00}|^2},
\end{equation}
where ${\cal G}_{00}\equiv{\cal G}_{00}(E\pm i0^+,0)$. 
We see that, for both choices 
of the sign, the position of the pole is shifted onto the unphysical sheet. 
The resonance width is second-order in $t/g_1$, while the width of the 
boson band (\ref{bandwidth}) is first-order. Therefore, in the limit of 
a small $t/g_1$, resonance bosons can propagate during their lifetime. 

\section{Conclusion}\label{conclusion}

We have discussed the properties of extended two-fermion bound states and 
resonances on a lattice by using an idealized Hubbard-like model with 
finite-range interactions.  This two-body problem is exactly
solvable on any periodic lattice, so that all approximations are well 
controlled.  The two-body states may be relevant to high-temperature 
superconductors for two reasons.  First, when the size of a 
Cooper pair in a superconductor 
is comparable to the average inter-particle distance, a two-fermion
picture may provide a better starting point than a Fermi-sea description.  
The superconducting transition would then be similar to the Bose 
condensation in liquid $^4$He. 
Second, any description of $d$-wave superconductivity must go beyond
a zero-range interaction between fermions --- such as in the BCS theory 
--- simply because the $d$-wave order parameter vanishes for two fermions on 
the same site.  

The symmetry of the order parameter $\Delta({\bf r},{\bf r}')$ in our picture 
is related to that of the ground state of a pair.  This observation
pinpoints the important role of the fermion kinetic energy in 
determining the symmetry of the energy gap.  It appears to us that 
the $d$-wave symmetry is related to the reversed sign of the hopping 
amplitudes along the diagonals of Cu--O sheets, a consequence
of the $d_{x^2-y^2}$ character of doped holes on copper sites.  We have
demonstrated that, when the magnitude of such an amplitude is comparable 
to that for nearest neighbors, the ground state of two fermions will 
be a $d_{x^2-y^2}$ wave.  

In the strong-binding case, we have shown that 
pairs formed by fermions on adjacent lattice sites can easily move 
through the lattice.  Thus, the condensation temperature need not 
be vanishingly low in this limit.  This observation resloves the 
large-effective-mass problem 
raised by Nozi\'eres and Schmitt-Rink in this connection \cite{Nozieres}.
Our conclusions in this regard are valid for narrow two-fermion 
resonances as well. 

For weakly bound fermions, we have found that the size of a 
pair having a non-trivial symmetry tends to be smaller than the
size of an $s$ wave.  Only in the latter case is the size determined 
by the depth of the bound-state energy level and given by the 
tunneling length $\xi_0=\hbar v_F/\Delta$.  In the former case, however, 
a pair's size is also sensitive to a short-range cutoff, which we have 
identified with the range of the attractive interaction.  This property 
has its origin in the weaker tunneling exhibited by states 
with higher angular momenta.  It is therefore not surprizing that 
the cuprates, quite possibly $d$-wave superconductors, have remarkably
small Cooper pairs. 

\section*{Acknowledgments}

The authors thank Professor T.~D.~Lee for encouragement and discussions. 
This work has been supported in part by U.S.~Department of Energy grants
DE-FG02-92 ER40699 and DOE-91 ER40651, Task B.

\appendix

\section{Nearest-neighbor attraction and many-body bound states}
\label{many-body}

To clarify the issue of many-body bound states raised in Section 
\ref{deep}, we consider the following simple model on a square lattice.  
In the limit of strong attraction the fermion kinetic energy can 
be neglected.  The interactions are given by an infinite on-site 
repulsion, an attractive potential $g_S<0$ for nearest neighbors 
in the spin-singlet state, and a repulsive potential $g_T>0$ for 
nearest neighbors in the 
spin-triplet states.  Then for two nearest neighbors, the Hamiltonian
can be written as
\begin{eqnarray}
H_{\rm n.n.} 
&=& \frac{3g_T+g_S}{4} + (g_T-g_S)({\bf S}_1\!\cdot\!{\bf S}_2)
\nonumber
\\
&=& \frac{g_T+g_S}{2} + \frac{g_T-g_S}{2}{\rm P}_{12},
\end{eqnarray}
where ${\bf S}_1$ and ${\bf S}_2$ are fermionic spins and ${\rm P}_{12}$ 
is the exchange operator. 

Three fermions on a square lattice will lower their energy to 
\begin{equation}
E_{2+1} = g_S<0
\end{equation}
by forming a pair and leaving one fermion free.  
Alternatively, there may be two bonds connecting one fermion 
with the other two.  The lowest energy of this system will be 
given by the smallest eigenvalue of the Hamiltonian,
\begin{equation}
H_3 = \frac{g_S+3g_T}{2} 
+ (g_T-g_S){\bf S}_1\!\cdot\!({\bf S}_2+{\bf S}_3),
\end{equation}
which is 
\begin{equation}
\label{E_3}
E_3 = \frac{3g_S+g_T}{2}.
\end{equation}
The ``2+1'' configuration then has a lower energy if $g_T>-g_S>0$.  
This condition can be relaxed in a fermionic ensemble.  Indeed, 
if there are two sets of ``2+1'' clusters on the lattice, the 
two free fermions will pair and reduce the energy further.  
Therefore, in a large ensemble, 
\begin{equation}
E_{2+1} = \frac{3g_S}{2},
\end{equation}
and the stability of pairs requires that $g_T>0$, {\em i.e.} the 
triplet interaction must be repulsive. 

\begin{figure}
\begin{center}
\epsfxsize=3.125in
\leavevmode
\epsffile{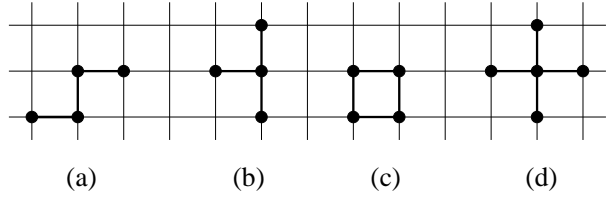}
\end{center}
\caption{Bond clusters on a square lattice.}
\label{clusters}
\end{figure}

Four fermions have a larger number of states from which to choose.  
The paired state ``2+2'' has the lowest energy 
\begin{equation}
\label{2+2}
E_{2+2} = 2g_S.
\end{equation}
Other possibilities include clusters of four spins having three 
or four bonds [Figures \ref{clusters}(a), (b), and (c)] with the following 
energies:
\begin{mathletters}
\begin{eqnarray}
\label{4}
E_4^{(a)} &=& \frac{(3+\sqrt{3})g_S + (3-\sqrt{3})g_T}{2},\\
E_4^{(b)} &=& 2g_S+g_T,\\
E_4^{(c)} &=& 3g_S+g_T.
\end{eqnarray}
\end{mathletters}
A comparison between Equations (\ref{2+2}) and (\ref{4}, b, and c) shows that 
pairs will 
merge into squares unless there is a strong enough triplet repulsion $g_T$.
We thus obtain a condition  necessary to prevent the many-body 
clustering in the limit of strong attraction:
\begin{equation}
\label{necessary}
g_T>-g_S.
\end{equation}

Extending the analysis in this form to larger clusters of fermions is 
impractical due to an exponentially growing number of 
possible ground states.  Instead, we shall give an upper bound for 
the triplet repulsion that ensures the stability of pairs against the  
formation of many-body bound states.  
The idea is based on the following observation.  The ground-state 
energy of a fermionic cluster with $N$ bonds is obtained by minimizing 
the expectation value of the Hamiltonian
\begin{equation}
H = 
\sum_{\langle ij\rangle}\left\{
\frac{3g_T+g_S}{4} \ +\ 
(g_T-g_S)({\bf S}_i\!\cdot\!{\bf S}_j)\right\}.
\end{equation}
Minimizing the energy of each bond separately gives a crude 
estimate of this energy as $Ng_S$, obviously much lower than
it is.  One can do better, however, by minimizing the 
energies of separate small clusters of {\em bonds} (not 
spins).  Indeed, the lowest energy of a small bond cluster can
also be found by minimization over all states of the large spin 
system.  It is clear then that such a separate minimization 
results in a lower energy.  If the bonds are grouped in $k$ 
small clusters, we have
\begin{equation}
H = H_1 + H_2 +\ldots + H_k
\end{equation}
and therefore
\begin{equation}
\min\langle H\rangle 
\geq \min\langle H_1\rangle
+ \min\langle H_2\rangle
+\ldots+ \min\langle H_k\rangle.
\end{equation}
The equality holds if all minima are reached in the same state
of the spin system.  In general, however, $[H_i,H_j]\neq 0$ 
(whenever clusters $i$ and $j$ have common spins) and thus the 
eigenstates of $H_i$ and $H_j$ do not coincide.  
This consideration allows one to obtain a reasonable lower 
bound for the ground-state energy of any cluster having an even
number of bonds.  It can be shown that the Hamiltonian is a 
sum over pairs of bonds (with the two bonds of any pair 
sharing a common spin).  The ground-state energy of two such
bonds is $E_3$ given above Equation (\ref{E_3})].  Therefore, 
the energy of a cluster with $N_b$ bonds satisfies the inequality
\begin{equation}
\label{E_N_b} 
E_{N_b} > N_b\frac{3g_S+g_T}{4}.
\end{equation}
In the dimerized state, $N_s/2$ spin pairs will have the energy 
\begin{equation}
\label{E_N_s}
E_{N_s} = N_s\frac{g_S}{2},
\end{equation}
which should be lower than $E_{N_b}$ to prevent the many-body 
formation.  Replacing $E_{N_b}$ by its lower bound 
(\ref{E_N_b}), we find that
\begin{equation}
g_T > -g_S\left(3-\frac{2N_s}{N_b}\right)
\end{equation}
is sufficient for the stability of pairs.  The lowest possible ratio
for $N_s/N_b$ on a square lattice is 1/2, achieved when all 
sites on a large lattice are occupied by spins.  Therefore, 
\begin{equation}
\label{sufficient}
g_T > -2g_S.
\end{equation}
provides a sufficient condition for a square lattice.

One can see, however, that this condition can be relaxed.  
As has been mentioned, the largest number of bonds 
originating from one spin is achieved when all spins have 
four nearest neighbors.  In this situation, 
one improves the lower bound for the energy by dividing 
the system into clusters with four bonds originating from 
one spin [Figure \ref{clusters}(d)].  Thus, (\ref{sufficient}) 
can be replaced by a milder condition. 

A cluster with an odd number of bonds can 
be divided into a cluster with an even number of bonds and 
another one with three bonds.  Then one can readily verify that 
condition (\ref{sufficient}) again ensures the stability
of pairs.  

\section{Point groups of square and cubic lattices}
\label{symmetry appendix}

The point group of a square lattice consists of eight elements: a unit 
element; three proper rotations about the center of the square (90, 180, 270 
degrees); two reflections with respect to the coordinate axes
that are parallel to the square's sides and that have the square's center
as the origin; two reflections 
with respect to the two diagonals of the square. These eight elements are 
isomorphic to the group of matrices $\pm I_2, \pm \tau_1, \pm i\tau_2$ and 
$\pm \tau_3$ with $I_2$ the $2\times 2$ unit matrix and $\tau_j$ the Pauli 
matrices. There are five conjugate classes, which implies the following five 
irreducible representations: a unit representation $A_1$; 
three non-trivial one-dimensional representations $A_2$, 
$B_1$, and $B_2$, implemented by the functions $x_1y_2-x_2y_1$, $x^2-y^2$,
and $xy$, respectively; a two-dimensional representation $E$, transforming as 
the pair of functions $(x,y)$.

The proper cubic group consists of 24 elements. In addition to the unit 
element, there are three $C_4$ axes 
running through the centers of opposite faces of the cube (9 rotations); 
four $C_3$ axes along the diagonals (8 rotations); and 6 $C_2$ axes 
connecting opposite edges (6 rotations). These elements form five conjugate 
classes, which implies the existance of 
five irreducible representations: two one-dimensional representations $A_1$
and $A_2$, implemented by the functions 1 and $xyz$; a two-dimensional
representation $E$ transforming as the doublet $(x^2-y^2,\ 2z^2-x^2-y^2)$;
and two three-dimensional representations $T_1$ and $T_2$, implemented by
the triplets $(x,y,z)$ and $(yz,zx,xy)$. 

\section{The Green's Function at ${\bf P}=0$ for a Square Lattice}
\label{Elliptic appendix}

The two-particle Green's function was introduced in Section \ref{Schrodinger 
equation}. At a total Bloch momentum ${\bf P}=0$,  
\begin{equation}
\label{Green's defined, P=0}
G(E;{\bf r},{\bf r}')\equiv G(E,0;{\bf r},{\bf r}') 
 = \frac{1}{{\cal N}}\sum_{\bf q} \frac{\psi({\bf r}|{\bf q})\ 
\psi^*({\bf r}'|{\bf q})}{E-E_{\bf q}},
\end{equation}
where $\psi({\bf r}|{\bf q})=\sqrt{2}\cos{({\bf q}\!\cdot\!{\bf r})}$ 
for ${\bf r}\neq0$, $\psi(0|{\bf q})=1$, 
and $E_{\bf q}$ is the kinetic energy of two fermions with Bloch 
momenta ${\bf q}$ and $-{\bf q}$,
\begin{equation}
\label{Eq}
E_{\bf q} = 2\sum_{\bf r} t({\bf r})e^{i{\bf q}\cdot{\bf r}}.
\end{equation}
Introduce 
\begin{equation}
\label{psi introduced}
\psi(E;{\bf r}) = \frac{1}{{\cal N}}\sum_{\bf q} 
\frac{e^{i{\bf q}\cdot{\bf r}}}{E-E_{\bf q}},
\end{equation}
which is the Green's function for unsymmetrized states, with 
$\psi({\bf r}|{\bf q}) = 
e^{i{\bf q}\cdot{\bf r}}$. If ${\bf r}\neq 0$ and ${\bf r}'\neq 0$, 
\begin{eqnarray}
\label{G in terms of psi}
G(E;{\bf r},{\bf r}') &=& \psi(E;{\bf r}-{\bf r}') 
+ \psi(E;{\bf r}+{\bf r}'),\\
G(E;{\bf r},0) &=& G(E;0,{\bf r}) = \sqrt{2}\psi(E;{\bf r}),\\
G(E;0,0) &=& \psi(E;0).
\end{eqnarray}
It is straightforward to verify that $\psi(E;{\bf r})$ satisfies the equation
\begin{equation}
\label{Laplace}
E\psi(E;{\bf r}) - 2\sum_{{\bf r}'}t({\bf r}')\psi(E;{\bf r}+{\bf r}') 
= \delta_{{\bf r},0}.
\end{equation}

We shall restrict the hopping range to nearest neighbors (amplitude $t_1$) and 
next-nearest neighbors (amplitude $t_2$). Then the kinetic energy of 
two fermions is
\begin{equation}
E_{\bf q} = 4t_1(\cos{q_x a} + \cos{q_y a}) + 8t_2\cos{q_x a}\cos{q_y a}.
\end{equation}
The minimum of $E_{\bf q}$ appears at ${\bf q}=0$ when both $t_1<0$ 
and $t_1+2t_2<0$. 
As has been shown in section \ref{symmetry of the ground state}, the 
ground-state wave function of two fermions has an alternating sign 
only if there exist positive hopping amplitudes. We therefore consider 
the case $t_2>0$, which may be relevant to cuprate superconductors.

At ${\bf P}=0$, the two-fermion energy band lies between $E_{\rm min} = 
E_{(0,0)} = 8t_1+8t_2$ and $E_{\rm max}= E_{(\pi,\pi)} = -8t_1+8t_2$. 
The density of states diverges when the line of constant $E$ touches 
the boundary of the Brillouin zone, at $E_{\rm vH}=E_{(\pi,0)} = -8t_2$. 
It will be convenient to measure the energy from the van Hove singularity: 
\begin{eqnarray}
\label{E'}
E' &=& E + 8t_2,\\
E'_{\rm min} &=& 8t_1+16t_2\leq 0,\\
E'_{\rm max}&=& -8t_1+16t_2 > 0.
\end{eqnarray}
On an infinite lattice, the Green's function for 
relative separation ${\bf r}=(m,n)$ is 
\begin{eqnarray}
\label{double int}
&&\psi(m,n) = \\
&&\int\frac{d^2{\bf q}}{(2\pi)^2}
\frac{\cos{mq_x}\cos{nq_y}}
{E-4t_1(\cos{q_x}+\cos{q_y})-8t_2\cos{q_x}\cos{q_y}}.
\nonumber
\end{eqnarray}
When $n=0$, the integration over $q_y$ is particularly simple. We have 
\begin{eqnarray}
\label{psi(m,0)}
&&\psi(m,0) = \\
&&\frac{1}{\pi}\int_0^\pi \frac{\cos{mq_x}\ dq_x}
{\sqrt{[E'-E'_{\rm min}\cos^2{(q_x/2)}][E'-E'_{\rm max}\sin^2{(q_x/2)}]}}.
\nonumber
\end{eqnarray}
The sign of the square root is determined as follows.  When the root 
is real, it has the same sign as $E'$, which ensures 
that $\psi(E;0,0)\approx 1/E$ as $E\to \infty$.  The region for 
imaginary values 
of the square root, between the branch points 
$E'_{\rm min}\sin^2{(q_x/2)}$ and 
$E'_{\rm max}\cos^2{(q_x/2)}$,
is accessed either from $E'=E'_{\rm max}\cos^2{(q_x/2)}+0^+>0$ 
(by encircling the singularity counterclockwise), 
or from $E'=E'_{\rm min}\sin^2{(q_x/2)}-0^+<0$ (by encircling the singularity 
clockwise). In both cases, 
the square root has a positive imaginary part and zero real part.

\begin{figure}
\begin{center}
\input{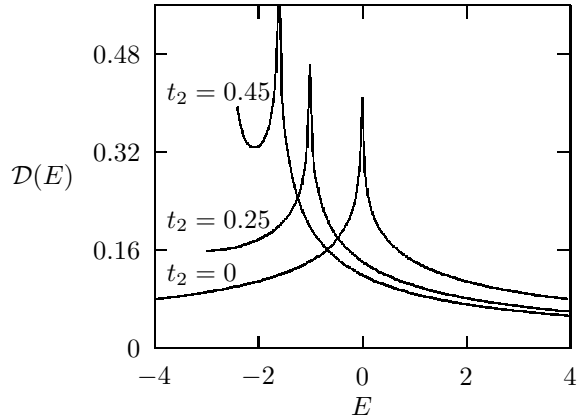}
\end{center}
\caption{Density of states for different values of nearest-neighbor hopping
$t_2$.  $t_1=-1$.  Values of $t_2$ are shown on the graph.}
\label{DOS plots}
\end{figure}

At the origin, $\psi$ can be expressed in terms of a complete elliptic integral
of the first kind,
\begin{equation}
\label{psi(0,0) outside}
\psi(0,0) = \frac{2}{\pi E'}{\bf K}(k),
\end{equation}
where 
\begin{equation}
\label{modulus}
k^2 = 1-(1-E'_{\rm min}/E')(1-E'_{\rm max}/E').
\end{equation}
This function can be analytically continued into the region 
$E'_{\rm min}<E'<E'_{\rm max}$ using the identity 
\begin{equation}
\label{continuation of K}
{\bf K}(k) = \frac{{\bf K}(1/k) \pm i{\bf K}'(1/k)}{k}
\end{equation}
(the sign is chosen in accordance with the rule stated above).
The density of states is then given by
\begin{equation}
\label{DOS}
{\cal D}(E) = -\frac{{\rm Im} G(0,0)}{\pi} 
= \frac{2}{\pi^2 |E'|k}{\bf K}'(1/k).
\end{equation}
This function is plotted in Figure \ref{DOS plots} for $t_1=-1$ and several 
positive values of $t_2$. 
As $t_2\to -t_1/2$, the band flattens at the bottom and the logarithmic van 
Hove singularity coalesces into the branch point of the Green's function.   
This strongly enhances the density of states near the lower edge of the band. 

As $m$ increases, expressions for $\psi(E;m,0)$ become progressively more 
complicated and include, in general, complete elliptic integrals of all 
three kinds. For example, 
\begin{eqnarray}
\psi(1,0) &=& 
\frac{4}{\pi}\left(\frac{1}{2E'}-\frac{1}{E'_{\rm max}}\right)
{\bf K}(k)\nonumber\\
\label{psi(1,0)}
&&-\frac{4}{\pi}\left(\frac{1}{E'}-\frac{1}{E'_{\rm max}}\right)
\Pi(E'/E'_{\rm max},k),
\end{eqnarray}
where the integral of the third kind is defined as 
\begin{equation}
\label{third kind}
\Pi(\alpha^2,k) = \int_0^1\frac{dx}{(1-\alpha^2x^2)\sqrt{(1-x^2)(1-k^2x^2)}}.
\end{equation}
An expression like (\ref{psi(1,0)}) provides little insight into 
the behavior of the Green's function. 
Fortunately, it is possible to obtain an approximation for
 $\psi(E;{\bf r})$ in the 
vicinity of its singular points. Several authors \cite{Dagotto} 
have identified the enhanced 
density of states near the van Hove singularity as a possible candidate for 
explaining high critical temperatures in cuprate superconductors. In view of 
that, we discuss the leading behavior of $\psi(E;{\bf r})$ near 
$E=E_{\rm vH}$, {\em i.e.} $E'=0$.

As $E'\to 0$, $k\to \infty$, so that ${\bf K}(1/k)\to \pi/2$ and 
\mbox{${\bf K}'(1/k)-\ln{4k}\to 0$}.
Therefore, near $E'=0$,
\begin{equation}
\label{psi(0,0) near vH}
\psi(0,0) \approx \frac{1}{E'_0} - \frac{2i}{\pi E'_0}\log{\frac{4E'_0}{E'}}.
\end{equation}
Note that ${\rm Re}{\psi(0,0)}\to \pm1/E'_0$ as $E'\to 0^\pm$, 
in accordance with 
the sign rule for the square root.
At even sites along the axes, the difference $\psi(2m,0)-\psi(0,0)$ 
remains finite as $E'\to 0$: 
\begin{eqnarray}
\psi(2m,0)-\psi(0,0) 
&\to& \frac{2i}{\pi E'_0} \int_0^\pi dq_x 
\frac{1-\cos{2mq_x}}{\sin{q_x}} \nonumber\\
\label{psi(2m,0)-psi(0,0)}
&=& \frac{8i}{\pi E'_0}\sum_{j=1}^{m}\frac{1}{2j-1}.
\end{eqnarray}
In a similar way, 
\begin{equation}
\label{psi(2m+1,0)-psi(1,0)}
\psi(2m+1,0)-\psi(1,0)\to 0.
\end{equation}
It is also seen from (\ref{psi(m,0)}) that the two divergent contributions 
to  $\psi(2m+1,0)$, near $q_x=0$ and $q_x=\pi$, have opposite signs and 
thus tend to cancel each other. In the limit as $E'\to 0$, 
\begin{equation}
\label{psi(1,0) approx}
\psi(1,0)\to 
\frac{1}{E'_0}-\frac{i}{\pi E'_0}\log{\frac{E'_{\rm max}}{-E'_{\rm min}}}. 
\end{equation}
This shows that $\psi(2m+1,0)$ is not singular at $E=E_{\rm vH}$. 

Using Equation (\ref{Laplace}) it is now possible to obtain approximations for 
$\psi({\bf r})$ on all sites. We demonstrate how this is done for the singular
part of the Green's function. As $E \to E_{\rm vH} = -8t_2$, 
Equation (\ref{Laplace}) reads 
\begin{equation}
\label{Laplace at van Hove}
-8t_2\psi({\bf r}) - 2t_1\sum_{\rm n.n.}\psi({\bf r}') 
- 2t_2\sum_{\rm n.n.n.}\psi({\bf r}'') 
= {\cal O}(1),
\end{equation}
where ${\bf r}'$ refers to the four nearest neighbors of ${\bf r}$,
while ${\bf r}''$ refers to the four next 
nearest sites. Starting with ${\bf r}=0$, we obtain (by using symmetry of the 
lattice) 
\begin{equation}
\label{psi(1,1) leading}
\psi(1,1) = - \psi(0,0) + {\cal O}(1).
\end{equation}
Applying Equation (\ref{Laplace at van Hove}) to site $(1,0)$, we find that  
\begin{equation}
\label{psi(2,1)}
\psi(2,1) = {\cal O}(1),
\end{equation}
because the divergent contributions of sites $(0,0)$ and $(2,0)$ are cancelled 
by those of $(1,1)$ and $(1,-1)$.
We now separate the lattice into two sublattices, putting adjacent sites  
on different sublattices.
In general, $\psi({\bf r})$ is singular on the sublattice that contains the 
origin: at sites with two even coordinates, 
$\psi({\bf r}) = \psi(0,0) + {\cal O}(1)$; at sites with two odd coordinates, 
$\psi({\bf r}) = -\psi(0,0) + {\cal O}(1)$. On the other sublattice, 
$\psi({\bf r})$ is 
regular at $E=E_{\rm vH}$. This means that the relative motion of the 
particles at $E=E_{\rm vH}$ is restricted to one of the sublattices. 

As we mentioned, on any site ${\bf r}$, $\psi({\bf r})$ is a linear 
combination of three 
complete elliptic integral with modulus (\ref{modulus}). Therefore, it should
be possible to find $\psi({\bf r})$ on the whole lattice 
if it is known on three 
conveniently chosen sites. For example, one starts out with (0,0), (1,0), and 
(2,0). Then $\psi(1,1)$ is found from the Laplace equation (\ref{Laplace}) 
around site (0,0) by using the symmetry arguments $\psi(m,n)=\psi(n,m)$ 
and $\psi(-m,n)=\psi(m,n)$. The Laplace equation on 
site (1,0) expresses $\psi(2,1)$ in terms of $\psi$ at sites (0,0), (1,0), 
(1,1), and (2,0). To proceed, we use an 
idea due to Friedberg and Martin \cite{FM}, who showed that there is a 
lattice analog 
of the gradient condition, $\nabla\psi({\bf r}) ||\ {\bf r}$. 
Extending their idea 
to the case of several hopping amplitudes, we find an additional relation 
between values of $\psi({\bf r}+{\bf r}')$ on sites surrounding ${\bf r}$.
\begin{equation}
\label{gradient}
{\sum_{{\bf r}'}}^\prime\ \frac{t({\bf r}')}{x'}\psi({\bf r}+{\bf r}') 
= {\sum_{{\bf r}'}}^\prime\ \frac{t({\bf r}')}{y'}\psi({\bf r}+{\bf r}'),
\end{equation}
where $\sum^\prime$ means that sites with $x'=0$ ($y'=0$) are excluded. 
On sites with high symmetry, such as $(m,m)$, $(m,0)$, and $(0,m)$, 
this equation becomes an identity and supplies no additional information. 

Now we are in a position to recover $\psi$ on the remainder of the lattice. 
We set up the Laplace equations on sites (2,0) and (2,1) in order to relate 
$\psi$ at 
sites (3,0), (3,1), and (3,2) to each other and to $\psi$ at sites where 
it has been 
calculated. The gradient equation on site (2,1) becomes a third one 
in this set of linear equations, which yields three more sites. $\psi(3,3)$ 
is obtained from the Laplace equation on site (2,2). This establishes the 
algorithm for computing $\psi$ on the remaining sites. When $\psi$ is known 
on the square with the corners $(0,0),\ (0,m),\ (m,0)$, and $(m,m)$, 
we write the 
Laplace equation for site $(m,0)$ and the Laplace and gradient equations for 
site $(m,1)$.  We then
solve these equations for $\psi$ at $(m+1,0),\ (m+1,1)$, and $(m+1,2)$. 
Using the Laplace 
equations for sites $(m,2),\ldots,(m,m)$, we move up and complete the 
side of the square. 

In two particular cases, $t_2=0$ and $2t_2=-t_1$, the calculation of 
$\psi({\bf r})$
is simplified by the absence of elliptic integrals of the third kind. In the 
latter case, the van Hove singularity coalesces with the bottom of the 
continuum band, $E'_{\rm min}=0$, resulting in a stronger divergence of 
$\psi(E;{\bf r})$ near $E'=0$. We write down an approximate expression for 
$\psi(E,{\bf r})$ without presenting a derivation. Just outside the continuum 
band, as $E'\to 0^-$,
\begin{eqnarray}
\label{psi(0-;m,n)}
&&\psi(m,n) \sim 
-\frac{4|mn|}{E'_{\rm max}}
-\frac{1}{\pi\sqrt{-E'E'_{\rm max}}}\\
&&\times\left[\log{\left(-\frac{16E'_{\rm max}}{E'}\right)} 
- 4\sum_{k=1}^m\frac{1}{2k-1} - 4\sum_{l=1}^n\frac{1}{2l-1}\right]
,\nonumber
\end{eqnarray}
which can be readily continued to the region $0<E'<E'_{\rm max}$; 
the point $E'=0$
should be encircled clockwise.

\section{Proof of identities used in Section V}
\label{resonance appendix}

To prove the relation
\begin{equation}
\label{-1}
-\Delta^\dagger \frac{\partial {\cal G}}{\partial\! E}\Delta = 1,
\end{equation} 
we write the quantity 
$-\Delta^\dagger (\partial {\cal G}/\partial\! E)\Delta\equiv\delta$ in its 
explicit form,
\begin{mathletters}
\begin{eqnarray}
\label{Delta dG/dE Delta}
\delta = -\sum_{i,j=0}^n\psi^*({\bf R}_i)g({\bf R}_i)
\frac{\partial G({\bf R}_i,{\bf R}_j)}{\partial\! E}
g({\bf R}_j)\psi({\bf R}_j).
\end{eqnarray}
It follows then from the definition of the Green's function 
(\ref{Green's defined}) and the orthogonality relation 
(\ref{orthonormal}) for functions $\psi({\bf r}|{\bf p})$ that 
\begin{eqnarray}
\frac{\partial G({\bf r}',{\bf r})}{\partial\! E} = 
-\sum_{{\bf r}''}G({\bf r}',{\bf r}'')G({\bf r}'',{\bf r}').\nonumber
\end{eqnarray}
The sum must be taken over the entire lattice, so that this identity 
cannot be applied directly to the $n\!\times\! n$ matrix 
${\cal G}(E,{\bf P})$.
We substitute this expression into (\ref{Delta dG/dE Delta}):  
\begin{eqnarray}
\delta &=& \sum_{i,j=0}^n\sum_{{\bf r}} 
\psi^*({\bf R}_i)g({\bf R}_i)G({\bf R}_i,{\bf r})G({\bf r},
{\bf R}_j)g({\bf R}_j)\psi({\bf R}_j)\nonumber \\
\label{sum extended}
&=& \sum_{{\bf r},{\bf r}',{\bf r}''}
\psi^*({\bf r}'')g({\bf r}'')G({\bf r}'',{\bf r})G({\bf r},
{\bf r}')g({\bf r}')\psi({\bf r}') \\
\label{use eqn of mtn}
&=& \sum_{{\bf r}}\psi^*({\bf r})\psi({\bf r}) = 1.
\end{eqnarray}
\end{mathletters}
In (\ref{sum extended}), the newly added terms do not contribute because
either $\psi({\bf r}')=0$ or $g({\bf r}')=0$ for sites 
${\bf r}'\neq{\bf R}_j$. The transition from (\ref{sum extended}) to
(\ref{use eqn of mtn}) is justified by Equations 
(\ref{psi = G g psi near cut}) and (\ref{psi* = psi* g G near cut}).

Next, we prove that, in ${\cal A}\,(\partial {\cal G}/\partial\! E)\Delta$, 
$\partial {\cal G}/\partial E$ can be replaced by $-{\cal G}^2$. Explicitly, 
\begin{eqnarray}
&&\sum_{j,k=0}^n A({\bf R}_i,{\bf R}_j)\frac{\partial G({\bf R}_j,{\bf R}_k)}
{\partial\! E}g({\bf R}_k)\psi({\bf R}_k) \nonumber \\ 
&=& -\sum_{j,k=0}^n \sum_{\bf r} A({\bf R}_i,{\bf R}_j)G({\bf R}_j,{\bf r})
G({\bf r},{\bf R}_k)g({\bf R}_k)\psi({\bf R}_k) \nonumber \\
&=& -\sum_{j=0}^n \sum_{\bf r} A({\bf R}_i,{\bf R}_j)
G({\bf R}_j,{\bf r})\psi({\bf r}) \nonumber\\ 
&=& -\sum_{j,k=0}^nA({\bf R}_i,{\bf R}_j)G({\bf R}_j,{\bf R}_k)\psi({\bf R}_k) 
\equiv -{\cal A}{\cal G}\psi.
\end{eqnarray}

\end{document}